\documentclass[12pt,eqsecnum]{article}
\usepackage{amsmath,amssymb,graphicx} 

\usepackage{psfrag}
\usepackage{subfigure}
\usepackage{tabularx}

\parskip=\baselineskip
\renewcommand{\baselinestretch}{1.2}

\newcommand{\ZZ}{\mathbb{Z}}

\newcommand{\myfig}[3]{
	\begin{figure}[ht]
	\centering
	\includegraphics[width=#2cm]{#1}\caption{#3}\label{fig:#1}
	\end{figure}
	}

\newcommand\cc[1]{#1^{^{\kern-6pt \circ}}\kern2pt}

\def\pa{\partial}
\renewcommand{\a}{\alpha}

\def\be{\begin{equation}}
\def\ee{\end{equation}}
\def\bea{\begin{eqnarray}}
\def\eea{\end{eqnarray}}
\def\ba{\begin{array}}
\def\ea{\end{array}}
\def\bi{\begin{itemize}}
\def\ei{\end{itemize}}

\def\Tr{{\rm Tr}}

\newcommand{\beq}{\begin{equation}}
\newcommand{\eeq}{\end{equation}}
\newcommand{\beqn}{\begin{eqnarray}}
\newcommand{\eeqn}{\end{eqnarray}}
\newcommand{\bga}{\begin{align}}
\newcommand{\w}{\wedge}

\newcommand{\h}{\hspace}
\def\dalemb#1#2{{\vbox{\hrule height .#2pt
\hbox{\vrule width.#2pt height#1pt \kern#1pt
\vrule width.#2pt}
\hrule height.#2pt}}}

\begin{document}

\def\thisaddress{\small Department of Physics, University of Illinois, 1110 W. Green St., 
Urbana IL 61801-3080, U.S.A. }
\def\thataddress{\small Department of Physics, University of Crete, Heraklion 71003, Greece}
\renewcommand\author[1]{#1}

\begin{center}
{\Large { Torsion and the Gravity Dual of Parity \\ Breaking  in AdS$_4$/CFT$_3$ Holography}}
\end{center}
\vskip 2 cm
\centerline{{\bf 
\author{Robert G. Leigh$^1$,}
 \author{Nam Nguyen Hoang$^1$}
and 
\author{Anastasios C. Petkou$^2$}}}

\vspace{.5cm}
\centerline{$^1$\it \thisaddress}
\vspace{.5cm}
\centerline{$^2$\it \thataddress}

\vspace{2cm}

\begin{abstract}
We study four dimensional gravity with a negative cosmological constant deformed by the Nieh-Yan torsional topological invariant with a spacetime-dependent coefficient. We find an exact solution of the Euclidean system, which we call the torsion vortex, having two asymptotic AdS$_4$ regimes  supported by a pseudoscalar with a kink profile. We propose that the torsion vortex is the holographic dual of a three dimensional system that exhibits distinct parity breaking vacua. The torsion vortex represents a (holographic) transition between these distinct vacua. We expect that from the boundary point of view, the torsion vortex represents a `domain wall' between the two distinct vacua.

From a bulk point of view, we point out an intriguing identification of the  parameters of the torsion vortex with those of an Abrikosov vortex in a Type I superconductor. Following the analogy, we find that external Kalb-Ramond flux then appears to support bubbles of flat spacetime within an asymptotically AdS geometry.

\end{abstract}

\pagebreak

\parskip= 2pt
\renewcommand{\baselinestretch}{.2}
\tableofcontents

\parskip=10pt
\renewcommand{\baselinestretch}{1.2}

\section{Introduction and summary of the results}

AdS$_4$/CFT$_3$ is currently emerging as a novel paradigm of holography  that has qualitatively different properties from the more familiar AdS$_5$/CFT$_4$ correspondence. Particularly intriguing is the recent accumulation of evidence that AdS$_4$/CFT$_3$ can be used to describe a plethora of phenomena in $2+1$ dimensional condensed matter systems, such as quantum criticality \cite{QuantCrit1,QuantCrit2}, Quantum Hall transitions \cite{QuantHall1,QuantHall2,QuantHall3,QuantHall4}, superconductivity \cite{Supercond1,Supercond2,Supercond3,Supercond4,Supercond5}, supefluidity \cite{Chrisnew,Basu}  and spontaneous symmetry breaking \cite{SSB1,SSB2,SSB3}.  Furthermore, AdS$_4$/CFT$_3$ is the appropriate setup to study the holographic consequences of generalized electric-magnetic duality of gravity and higher-spin gauge fields \cite{LP,deHPConf,MPT1,MPT2,deH}.

In the absence of an explicit AdS$_4$/CFT$_3$ correspondence example,\footnote{The recently suggested field theoretic models for M2 branes \cite{BL1,BL2,BL3,Gustavsson,Sangmin,ABJM} are important steps towards the understanding of the boundary side of AdS$_4$/CFT$_3$.} various toy models have been used to study its general qualitative aspects. One of the aims of the present work it to provide yet another model that can be used to unveil some salient and intriguing properties of AdS$_4$/CFT$_3$ holography.  However, this is not our only aim. We also wish to study here the relevance of torsion to four dimensional gravity from a holographic point of view.  The study of torsion is an interesting subject in itself that poses formal and phenomenological challenges.\footnote{See \cite{Shapiro,Zanelli,Freidel:2005sn} for recent reviews and \cite{Mercuri,Canfora} for other recent works.} In the context of a string theory description of gravity, torsion is omnipresent through antisymmetric tensor fields. AdS$_4$/CFT$_3$ provides the basic setup where four dimensional torsion can be holographically investigated. 

We consider a simple toy model where torsion is introduced via the topological Nieh-Yan class. In particular, we consider the modification of the Einstein-Hilbert action with a negative cosmological constant by the Nieh-Yan class, the latter having a spacetime-dependent coefficient. In the context of the 3+1-split formalism for gravity \cite{LP} we point out that the torsional degrees of freedom are carried by the `gravitational magnetic field.' In pure gravity the magnetic field is fully determined by the frame field, and torsion vanishes. In our model, the spacetime dependence of the Nieh-Yan coefficient makes some of the components of the magnetic field dynamical and as a consequence torsional degrees of freedom enter the theory. Our toy model is simple enough such that only one of the torsional degrees of freedom becomes dynamical. This degree of freedom can be either carried by a pseudoscalar, in which case our model is equivalent to a massless pseudoscalar coupled to gravity, or by a two-form gauge potential. In the latter case our model becomes equivalent to a Kalb-Ramond field coupled to gravity.  

Next, we find an exact solution of the equations of motion in Euclidean signature. Our metric ansatz is that of a domain wall (in the bulk). The solution, the {\it torsion vortex}, has two distinct asymptotically AdS$_4$ regimes along the ``radial" coordinate. The pseudoscalar has a kink profile and it is finite at both of the asymptotic regimes.  Our torsion vortex can be viewed as a generalization of the axionic wormhole solution of \cite{GS} in the case of non-zero cosmological constant. See also  \cite{Gutperle} for recent work on AdS wormholes. Having in mind the holographic interpretation of our model we focus mainly on the case where the torsional degree of freedom is carried by a pseudoscalar field. Following standard holographic recipes we find that the torsion vortex is the gravity dual of a three dimensional system that possesses two distinct parity breaking vacua. The two vacua are distinguished by the relative sign of the  pseudoscalar order parameter. 
Our bulk picture suggests that the transition from one vacuum to the other can be done by a marginal deformation of the theory. In Appendix \ref{app:C} we suggest that the above qualitative properties can be realized in the boundary by the three dimensional Gross-Neveu model coupled to U(1) gauge fields. 

Finally, we point out that the bulk physics of our vortex solution bears some resemblance to the Abrikosov vortex of superconducting systems. There is a natural mapping of the parameters of the torsion vortex to those of the Abrikosov vortex. 
We show that the gravitational parameter that is interpreted as an order parameter satisfies a $\phi^4$-like equation and this motivates us to suggest that the cosmological constant is related to the ``critical temperature" as  $\Lambda \sim T-T_c$. We end with a discussion of multi-vortex configurations and vortex condensation.    The outcome of this analysis is that $H$-flux supports bubbles of flat spacetime.

The paper is organized as follows. In Section 2 we discuss the relevance of torsional degrees of freedom in gravity and their relation with the gravitational magnetic field. In Section 3 we present our toy model and its various equivalent manifestations and discuss its 3+1-split formalism of \cite{LP}. 
In Section 4 we present the explicit torsion vortex solution of our model. In Section 5 we discuss the holography of the torsion vortex.
Section 6 contains the bulk physics of the vortex and its relationship to the Abrikosov vortex. It also contains the discussion of multi-vortices and vortex condensation. 
Technical details and the discussion of the three-dimensional Gross-Neveu model coupled to U(1) gauge fields are contained in the Appendices. 

\section{Torsional degrees of freedom in gravity}

\subsection{Preliminaries}

In this paper we will consider a four dimensional spacetime with a negative cosmological constant. The Einstein-Hilbert action may be written as\footnote{We use $I_{EH}=-16\pi G_4 S_{EH}$ where $S_{EH}$ is the usually normalized gravitational action. To fix notation we note that the Einstein equations that follow from $S_{EH}$ are $G_{\mu\nu}+\Lambda g_{\mu\nu}=0$. We will also write $\Lambda=-3\sigma_\perp/L^2$. }
\beq
I_{EH}=\int_M \left(\epsilon_{abcd} e^a \w e^b \w R^{cd} -  \frac{1}{6} \Lambda
\epsilon_{abcd} e^a \w e^b \w e^c \w e^d \right)\,,\label{actionEH}
\eeq
where $e^a$ denote the one-form frame fields, while ${\omega^a}_b$ are the connection one-forms with curvature ${R^a}_b=d{\omega^a}_b+{\omega^a}_c\w {\omega^c}_b$. As is well-known the variation of (\ref{actionEH}) gives the Einstein equations and also the zero torsion constraint $T^a=de^a+{\omega^a}_b\w e^b=0$. By virtue of the latter this action can be written entirely in terms of metric variables. 

There are also a number of other terms that one may consider. These are all of potential interest to holography  because being total derivatives  they may induce interesting boundary effects. We may parameterize these terms as follows (writing all possible $SO(3,1)$-invariant 4-forms constructed from $e^a, {R^a}_b, T^a$):
\beq
I_{top}=n\int_M C_{NY}+2\gamma^{-1}\int_M C_{Im}+p\int_M P_4+q\int_M E_4\,,
\eeq
where $C_{NY}= T^a\wedge T_a-R_{ab}\wedge e^a\wedge  e^b=d(T^a\wedge e_a)$ is the Nieh-Yan form, $\gamma$ is often referred to as the Immirzi parameter with $C_{Im}={R^a}_b\wedge e^b\wedge e_a$, $P_4=-\frac{1}{8\pi^2}{R^a}_b\w {R^b}_a=-\frac{1}{8\pi^2}d({\omega^a}_b\wedge {R^b}_a-\frac13 {\omega^a}_b\w {\omega^b}_c\w {\omega^c}_a)$ is the Pontryagin form and $E_4=-\frac{1}{32\pi^2}\epsilon_{abcd}R^{ab}\w R^{cd}$ is the Euler form. 
We note that $P_4+\frac{\sigma_\perp a^2}{4\pi^2} C_{NY}$ and $C_{NY}-C_{Im}$ are actually $SO(3,2)$ invariants \cite{Zanelli}.
These terms become of more interest, even in gravity, if we allow the coefficients to become fields. Although we will not consider this problem here in full generality, we will consider a particular example. We note that there is older literature, principally by d'Auria and Regge \cite{dr} that also considered some such cases (usually in asymptotically Minkowski geometries). In the course of the paper, we will review what is known from those older works. The purpose of our work, amongst other things, is to bring this up to date, and in particular focus on aspects of holography.


\subsection{Torsion and the magnetic field of gravity}

Our simple model involves only the Nieh-Yan (NY) term. It is interesting to discuss the physics of this topological invariant before we embark  on detailed calculations. We will see below that the NY term in gravity plays a role similar to that of the $\theta$-angle in gauge theories.

To see this, we explain below the relationship between the gravitational magnetic field $B^\alpha$ and torsion. Consider the 3+1 split\footnote{In appendix \ref{app:B} we present a brief review of the 3+1 split formalism where the definitions of the various relevant quantities appear and notation is explained. We note here that $\sigma$ is the overall signature of the spacetime, while $\sigma_\perp$ is the signature of the radial direction and $\sigma_3$ the signature of the boundary.} of the Einstein-Hilbert action (1) with the addition of the usual gravitational Gibbons-Hawking boundary term $I_{GH}$ \cite{LP,MPT1,MPT2}
\beqn
\label{3+1EH}
&&\hspace{-.5cm}I_{EH}+I_{GH} =\int dt\wedge\left(\dot{\tilde e}^\alpha\wedge (-4\sigma_\perp\epsilon_{\alpha\beta\gamma}\tilde{e}^\beta\wedge K^\gamma)\right.\hfill \nonumber \\
&&+2\sigma_\perp N\left\{ 2\tilde{d}\left(B^\alpha\wedge \tilde{e}_\alpha\right) +2B^\alpha\wedge\tilde{T}_\alpha +\epsilon_{\alpha\beta\gamma}\left( \sigma B^\alpha\wedge B^\beta-K^\alpha\wedge K^\beta-\frac{\sigma_\perp\Lambda}{3} {\tilde e}^\alpha\wedge {\tilde e}^\beta\right)\wedge {\tilde e}^\gamma 
\right\}\nonumber\\
&&-4\sigma_\perp{N}^\alpha\epsilon_{\alpha\beta\gamma}(\tilde D K)^\beta\wedge {\tilde e}^\gamma+4Q^\alpha(K_\beta \wedge\tilde e^\beta)\wedge\tilde e_\alpha+\left.4{q^0}_\alpha\left\{
{\epsilon^\alpha}_{\beta\gamma}\tilde{T}^\beta\wedge\tilde{e}^\gamma
\right\}\right)\,.
 \eeqn
In the 3+1 split formalism the dynamical variables in (\ref{3+1EH}) are the ``spatial"\footnote{By spatial, we will mean orthogonal to the ``radial" coordinate $t$. In the case of $AdS_4$, this radial coordinate is spacelike, and thus $\sigma_\perp=+1$.} one-forms $\tilde{e}^\alpha$, $K^\alpha$ and $B^\alpha$. The first two are canonically conjugate variables. The magnetic field $B^\alpha$ carries the torsional degrees of freedom as it can be seen for example if we write the definition of  the non-trivial `spatial' torsion as
\beq
\label{3torsion}
\tilde{T}^\alpha =\tilde{d}\tilde{e}^\alpha -\sigma\epsilon^{\alpha\beta\gamma}B_\beta\wedge\tilde{e}_\gamma\,.
\eeq
It is easily seen that the radial component of torsion $T^0$ is determined by $\tilde{e}^\alpha$ and $K^\alpha$. Notice that (\ref{3torsion}) implies that the tensor $B_{\alpha\beta}$ is odd under `spatial' parity, hence the trace $B^\alpha_{\,\,\alpha}$ is a pseudoscalar. Although a priori the torsional degrees of freedom are not connected with the pair of conjugate variables $\tilde{e}^\alpha $ and $K^\alpha$, they are not dynamical as there is no kinetic term for $B^\alpha$. Rather, they enter (\ref{3+1EH}) algebraically and as such they yield the algebraic zero torsion condition by virtue of which the magnetic field is related to the frame field. Indeed, as discussed in Ref. \cite{LP}, the $q^0_{\,\,\alpha}$ constraint sets to zero the antisymmetric part of $B^\alpha$ in deDonder gauge, such that the first term in the second line of (\ref{3+1EH}) vanishes. Then, the variation of (\ref{3+1EH}) with respect to $B^\alpha$ yields $\tilde{T}^\alpha=0$, leaving as true dynamical variables $\tilde{e}^\alpha$ and $K^\alpha$. This is the gravitational analogue of the electromagnetic case where the magnetic field is related to the gauge potential via the Bianchi identity. 

Consider now adding to the Einstein-Hilbert action the Nieh-Yan class $C_{NY}$ with a constant coefficient $\theta$. Over a compact manifold, the NY class is a topological invariant and takes integer values\footnote{More precisely, $C_{NY}/(2\pi L)^2$ is integral, as it is equal to the difference of two Pontryagin forms.} \cite{Zanelli}. Having in mind holography, we are interested here in manifolds with  boundary. In particular, the $3+1$ split has been set up so that the boundary is a constant-$t$ slice.  The NY term reduces to a boundary contribution. The explicit calculation yields
\beqn
\label{NY3+1}
{\cal I}_{NY} \equiv -2\sigma_\perp\theta\int C_{NY}
= 2\sigma_\perp \theta\int dt\wedge \left[2\epsilon_{\alpha\beta\gamma}\dot{\tilde{e}}^\a\wedge\tilde{e}^\beta\wedge B^\gamma +\epsilon_{\alpha\beta\gamma}\dot{B}^\alpha\wedge\tilde{e}^\beta\wedge\tilde{e}^\gamma\right]\,.
\eeqn
Adding (\ref{NY3+1}) to (\ref{3+1EH}) we obtain
\beqn
\label{3+1EHNY}
I_{EH}+I_{GH} +
{\cal I}_{NY}&=&\int dt\wedge\left(\dot{\tilde e}^\alpha\wedge (-4\sigma_\perp\epsilon_{\alpha\beta\gamma}\tilde{e}^\beta\wedge \left[ K^\gamma-\theta B^\gamma\right])+2\sigma_\perp\theta\epsilon_{\alpha\beta\gamma}\dot{B}^\alpha\wedge\tilde{e}^\beta\wedge\tilde{e}^\gamma \right.\hfill \nonumber \\
&&+\ {\rm constraint\ terms}\Big)\,.
 \eeqn
Notice that the ${\cal I}_{NY}$ term has two effects. One is to modify the canonical momentum variable $K^\alpha\mapsto K^\alpha -\theta B^\alpha$. This is analogous to the effect of the $\theta$-angle in the canonical description of electromagnetism \cite{Jackiw}.  The other is to 
provide a kinetic term for the singlet component of the magnetic field (one easily verifies that only ${B^\alpha}_{\alpha}$ contributes in the second term in the first line of (\ref{3+1EHNY})). This second effect has no analogue in electromagnetism. Taking the variation of (\ref{3+1EHNY}) with respect to $B^\alpha$, one finds that the zero torsion condition still holds.  This is expected of course since the ${\cal I}_{NY}$  term is  purely a boundary term. As a consequence, the true dynamical variables remain $\tilde{e}^\alpha$ and $K^\alpha$. However, the holography is slightly modified. The variation of (\ref{3+1EHNY}) gives on-shell 
\beq
\label{NYHol1}
\delta\left(I_{EH}+I_{GH}+{\cal I}_{NY}\right)_{on\,\,shell} = \int_{\partial {\cal M}}\delta\tilde{e}^\alpha\wedge\left(-4\sigma_\perp\epsilon_{\alpha\beta\gamma}\tilde{e}^\beta\wedge\left[K^\gamma-\theta B^\gamma\right]\right)_{on\,\,shell}\,.
\eeq
After the appropriate subtraction of divergences \cite{MPT1,MPT2}, (\ref{NYHol1}) yields a modified boundary energy momentum tensor. The modification is due to the term $4\sigma_\perp \theta\epsilon_{\alpha\beta\gamma}\tilde{e}^\beta\wedge B^\gamma$ which is parity odd and corresponds to the unique symmetric, conserved and traceless tensor of rank two and scaling dimension three that can be constructed from the three-dimensional metric \cite{LPsl2z}. It is  the exact analogue of the topological spin-1 current constructed from the three dimensional gauge potential.

The form of the action (\ref{3+1EHNY}) unveils an intriguing possibility. The above holographic interpretation was based on the zero torsion condition that connects $B^\alpha$ to the frame field. However, to get the zero torsion condition from (\ref{3+1EHNY}) we needed to integrate by parts the last term in the first line. Hence, if $\theta$ were $t$-dependent, the torsion would no longer be zero and the trace $B^\alpha_{\,\,\alpha}$ would become a proper dynamical degree of freedom independent of $\tilde{e}^\alpha$. In such a case the holographic interpretation of (\ref{3+1EHNY}) would change. The new bulk degree of freedom would couple to a new pseudoscalar boundary operator. As a consequence, we have the possibility to probe additional aspects of the boundary physics and describe new 2+1 dimensional phenomena. That we do in the next section. 

\section{The Nieh-Yan models}
\subsection{General aspects}

In the previous section we sketched a mechanism by which torsional degrees of freedom become dynamical. In particular, we have argued that the addition of the Nieh-Yan class with a space-time-dependent coefficient in the Einstein-Hilbert action makes dynamical a pseudoscalar degree of freedom connected to the trace of the gravitational magnetic field. 
Adding boundary terms to the bulk action corresponds to a canonical transformation. Consequently, by adding boundary terms we can change the canonical interpretation  and the variational principle.  Consider first the action
\beq\label{eq:INYprime}
I'_{NY}=I_{EH}[e,\omega]+I_{GH}[e,\omega]+2\int_M F(x) C_{NY}\,,
\eeq
where $F$ is a pseudoscalar `axion' field with no kinetic term. If $F\equiv -\sigma_\perp\theta$ were a constant, this theory would be equivalent to that studied in the last section. With $F=F(x)$, we have additional terms in the action involving gradients of $F$. If we perform the $3+1$ split on this action, we will find that $\tilde e^\alpha$ and $B^\alpha$ are canonical coordinates, and their conjugate momenta will depend on $F$.


 
The action as given may be supplemented by additional boundary terms. Such boundary terms are analogous to the Gibbons-Hawking term in gravity, but here involve the torsional degrees of freedom. In particular, we can replace $I'_{NY}$ by 
\beq
I_{NY}=I_{EH}[e,\omega]+I_{GH}[e,\omega]-2\int_M dF\w T_a\w e^a\,.
\eeq
This action is such that $\tilde e^\alpha$ and $F$ are canonical coordinates with appropriate boundary conditions, while $B^\alpha$ appears in the  momentum conjugate to $F$. 
To investigate this theory, we note that the variation of the action takes the form
\beqn
\delta I_{NY}=2\int_M \delta e^d\w  \left[\epsilon_{abcd} e^b\w\left(R^{cd}-\frac13\Lambda e^c\w e^d\right)+2dF\w T_d\right]\nonumber \\
+2\int_M\delta\omega^{ab}\w\left[ \epsilon_{abcd}T^c\w e^d+ dF\w e_b\w e_a
\right]+2\int_M\delta F\ C_{NY}\nonumber \\
+2\int_Md[\delta e^a\wedge\left(\epsilon_{abcd} e^b\w\omega^{cd}-dF\w e_a\right)-T_a\w e^a\delta F ]\,.
\eeqn
A non-trivial configuration of $F$ would source a particular component of the torsion. Indeed the classical equations of motion can be manipulated to yield in the bulk
\beq\label{eq:Ftors}
T^a\w e_a=3*_4 dF\,,
\eeq
where $*_4$ denotes the Hodge-$*$ operation.
However, as d'Auria and Regge \cite{dr} showed, this classical system is equivalent to a pseudoscalar
coupled to torsionless gravity. 
\beq\label{eq:PSL}
I_{PS}=I_{EH}[e,\cc{\omega}]+I_{GH}[e,\cc{\omega}]-3\int_M dF\w *_4 dF\,.
\eeq
This comes about as follows. We write the connection as $\omega=\cc{\omega}+\Omega$, where $\cc{\omega}$ is torsionless, and insert the
equation of motion (\ref{eq:Ftors}). The latter becomes an equation\footnote{Explicitly this is ${\Omega^a}_b=\frac{\sigma}{4}{\epsilon^{acd}}_b \pa_c F e_d$.} for $\Omega$, and we obtain (\ref{eq:PSL}). 

The holographic interpretation of a massless pseudoscalar field coupled to torsionless gravity is that it is dual to dimension $\Delta=3,0$ composite pseudoscalar operators in the boundary. The usual holographic dictionary then says that only the $\Delta=3$ operator appears in the boundary theory since only this is above the unitarity bound of the three dimensional conformal group $SO(3,2)$. A scalar operator with  dimension $\Delta=0$ would simply correspond to a constant in the boundary. Hence, the sensible holographic interpretation of the massless bulk pseudoscalar is that its leading behaviour determines the marginal coupling of a $\Delta=3$ operator; the expectation value of the operator itself is determined by the subleading behaviour of the bulk pseudoscalar. 

Another equivalent formulation of this bulk theory is obtained by writing
\beq
*_4dF={\frac{1}{3}} H\,.
\eeq
with $H$ a 3-form field.
 This is the parameterization that would be most familiar from string theory, 
as the system simply corresponds to an antisymmetric 2-form field. In this formulation, we write
\beqn
I_{KR}&=&I_{EH}[e,\cc{\omega}]+I_{GH}[e,\cc{\omega}]+\frac13\int_M H\w *_4H+\sqrt{\frac{2}{3}}\int_MC\w d*_4H\nonumber \\
&=&I_{EH}[e,\cc{\omega}]+I_{GH}[e,\cc{\omega}]-\frac12\int_M dC\wedge *_4dC+\int_M d(C\w *_4dC)\,.
\eeqn
In the first equation, $C$ appears as a Lagrange multiplier for the `Gauss constraint' and in the second expression, we have solved for the $H$ equation of motion in the bulk, which is just $H=\sqrt{\frac32}dC$.

\subsection{The 3+1-split of the pseudoscalar Nieh-Yan model}

To investigate the holographic aspects of our model it is most useful to use the `radial quantization' in which we think of the radial coordinate as `time' $t$. We have derived the radial 3+1 split in the first order
formalism in \cite{LP}, and this is summarized with explanations of notation in Appendix \ref{app:B}. Here we update that calculation to include torsional terms. The Nieh-Yan deformation gives
\beqn
-2\int dF\wedge T^a\wedge e_a &=& 2\int dt\w \Biggl\{-\dot F\tilde T_\alpha\w \tilde e^\alpha-\dot{\tilde e}_\alpha\w\tilde d F\w\tilde e^\alpha +N[2\tilde dF\w K_\alpha\w\tilde e^\alpha]\nonumber \\
 && \hspace{1.5cm}+N^\alpha[2\tilde dF\w\tilde T_\alpha]+Q^\alpha[-\sigma\epsilon_{\alpha\beta\gamma}\tilde dF\w\tilde e^\beta\w \tilde e^\gamma]\Biggl\}\,.
\eeqn
We see that the $F$ field makes a contribution to the constraints, and has a conjugate momentum proportional to the scalar part of the torsion (the part transverse to the radial direction).
The full bulk action becomes
\beqn
I&=&\int dt\wedge\left(\dot{\tilde e}^\alpha\wedge (4\sigma_\perp\epsilon_{\alpha\beta\gamma}K^\gamma\wedge\tilde e^\beta-2\tilde dF\wedge\tilde e_\alpha)-2\dot F (\tilde e^\alpha\wedge\tilde T_\alpha)\right.\nonumber \\
&&+N\left\{ 2\epsilon_{\alpha\beta\gamma}\left( ^{(3)}R^{\alpha\beta}-\sigma_\perp K^\alpha\wedge K^\beta-\frac{\Lambda}{3} {\tilde e}^\alpha\wedge {\tilde e}^\beta\right)\wedge {\tilde e}^\gamma 
+4\tilde dF\wedge K_\alpha\wedge\tilde e^\alpha
\right\}\nonumber\\
&&+4{N}^\alpha\left\{-\sigma_\perp \epsilon_{\alpha\beta\gamma}(\tilde D K)^\beta\wedge {\tilde e}^\gamma+\tilde dF\wedge\tilde T_\alpha
\right\}\nonumber \\
&&+4Q^\alpha\left\{ (K_\beta \wedge\tilde e^\beta)\wedge\tilde e_\alpha-\frac12\sigma\epsilon_{\alpha\beta\gamma}
\tilde dF\wedge\tilde e^\beta\wedge\tilde e^\gamma
\right\}\nonumber \\
&&+\left.4{q^0}_\alpha\left\{
{\epsilon^\alpha}_{\beta\gamma}\tilde{T}^\beta\wedge\tilde{e}^\gamma
\right\}\right)\,.
 \eeqn
We notice that the $Q$-constraint term can be written in the form
\beq\label{eq:Kcon}
4Q_\alpha\tilde e^\alpha\w\left( K_\beta \wedge\tilde e^\beta-\sigma *_3\tilde dF\right)\,.
\eeq
Because of this constraint (which relates the antisymmetric part of the extrinsic curvature to
the vorticity of $F$), the momentum conjugate to $\tilde e^\alpha$ is symmetric, i.e.
\beqn
\Pi_\alpha&=& 4\sigma_\perp\epsilon_{\alpha\beta\gamma}K^\gamma\wedge\tilde e^\beta-2\tilde dF\wedge\tilde e_\alpha\nonumber \\
&=& 4\sigma_\perp\left(\epsilon_{\alpha\beta\gamma}K^\gamma\wedge\tilde e^\beta-\frac12\sigma_3 *_3(K_\beta\w \tilde e^\beta)\w \tilde e_\alpha\right)\,.
\eeqn
When written out in components, one finds that the antisymmetric part $K_{[\alpha\beta]}$ cancels
\beq
\Pi_\alpha = 4\sigma_\perp (K_{(\beta\alpha)}-tr K\ \eta_{\beta\alpha})\tilde e^\beta\,.
\eeq
This result is consistent with the fact noted above, that the system may be equivalently described as
a pseudoscalar field coupled to torsionless gravity. Moreover, if  we take the deDonder gauge $d^\dagger\tilde e^\alpha=0$,  the torsion constraint implies that $B$ is symmetric. 

The ${q^0}_\alpha$ constraint yields ${\tilde T^\beta}_{\alpha\beta}=0$.
Out of the nine components of $\tilde T$, which transform as ${\bf 5}+{\bf 3}+{\bf 1}$ under $SO(3)$ (or $SO(2,1)$), this
sets the triplet to zero (the ${\bf 5}$ also vanishes on an equation of motion). The momentum conjugate to $F$ is given by
\beq
\Pi_F = -2\epsilon^{\alpha\beta\gamma}\tilde T_{\alpha\beta\gamma}\,.
\eeq
This is the singlet part of the torsion, which has become dynamical in this description of the theory, in the sense that it is canonically conjugate to $F$.

\section{The torsion vortex}

We will now simplify the analysis by taking a coordinate basis, and looking for solutions of the form
\beq
\tilde e^\alpha = e^{A(t)}dx^\alpha,\ \ \ \ N=1,\ \ \ \ N^\alpha=0\,,
\eeq
and we will further suppose that $F=F(t)$. In this case $K^\alpha$ and $B^\alpha$ reduce to one degree of freedom each as a result of the constraints
\beq
K_\alpha=k\tilde e_\alpha,\ \ \ \  B_\alpha=b\tilde e_\alpha\,,
\eeq
and one finds $\Pi_A=-4\sigma_\perp k$ and $\Pi_F=2\sigma b$. The action then takes the following relatively simple Hamiltonian form
\beq
I_{NY} \propto \int dt\,d^3x\,\,e^{3A(t)}\left[ \dot A\Pi_A+\dot F\Pi_F-\left(\frac12\sigma_3\Pi_F^2+\frac18\sigma_\perp\Pi_A^2+\frac23\Lambda\right)\right]\,.
\eeq
and the equations of motion give
\beqn
\label{eom1234}
&\dot\Pi_A=3\dot F\Pi_F,\ \ \ \ \ \ \dot\Pi_F+3\Pi_F\dot A=0,\ \ \ \ \ \
\Pi_A=4\sigma_\perp\dot A,\ \ \ \ \ \Pi_F=\sigma_3\dot F&\,,\\ 
\label{eom5}
&\Pi_A^2+4\sigma\Pi_F^2+\frac{16}{3}\sigma_\perp\Lambda=0&\,.
\eeqn
These equations of motion could of course alternatively  be obtained by considering
the theory in the form (\ref{eq:PSL}). It is convenient to rescale $F(t)=\frac13 \Theta(t)$. Then the equations of motion can be put in the form
\beq
\label{dwalleqs}
\ddot A+3\dot A^2-3a^2=0, \ \ \ \ \ \ \ddot A=\frac{1}{12}\sigma\dot \Theta^2, \ \ \ \ \ \ \ddot \Theta+3\dot \Theta\dot A=0\,.
\eeq
where we have set $\Lambda=-3\sigma_\perp a^2$ with $a=1/L$. These are of the standard form of domain wall equations that have appeared numerous times in the AdS/CFT literature. However, there is a crucial difference. Notice that the first two of (\ref{dwalleqs}) imply
\beq
\label{dwalleq1}
\dot{A}^2 +\frac{1}{36}\sigma\dot{\Theta}^2 -a^2=0\,.
\eeq
For Euclidean signature ($\sigma=\sigma_3=1$) the second term in (\ref{dwalleq1}) has {\it positive} sign in contrast to most of the other holographic studies. This is due to the fact that in passing from Lorentzian to Euclidean signature the pseudoscalar kinetic term acquires the `wrong sign' \cite{Gibbons}. This property allows for a remarkable exact solution to the above  system of non-linear equations in Euclidean signature, which we refer to as the {\it torsion vortex}.  To obtain it we define
\beq
h(t)=\dot A(t)\,,
\eeq
at which point we have
\beq\label{eq:hdot}
\dot h=\frac{1}{12}\dot \Theta^2,\ \ \ \dot h+3(h^2-a^2)=0\,.
\eeq
The general solution is of the form
\beq
h(t)=a\tanh 3a(t-t_0)
\eeq
and we then have
\beq
\label{Pf}
\Pi_F=\dot F=\pm2\sqrt{a^2-h^2(t)}=\pm 2a\ \text{sech}\ 3a(t-t_0)
\eeq
which gives
\beq
\label{Theta}
\Theta(t)=\Theta_0\pm 4\arctan\left(e^{3a(t-t_0)}\right)\,.
\eeq
The $\pm$ sign corresponds to kink/antikink and we will without loss of generality choose the $+$ sign.
We may also solve for
\beq\label{eq:expA}
e^{A(t)}=\alpha (2\cosh 3a(t-t_0))^{1/3}
\eeq
The parameter $\alpha$ is an arbitrary positive integration constant that sets the overall scale of the spatial part of the metric. $t_0$ may be interpreted as the position of the vortex; when $t_0=0$ the torsion vortex sits in the middle between the two asymptotically AdS$_4$ regimes.  
Below, we will discuss the interesting holographic interpretation of the torsion vortex.

\myfig{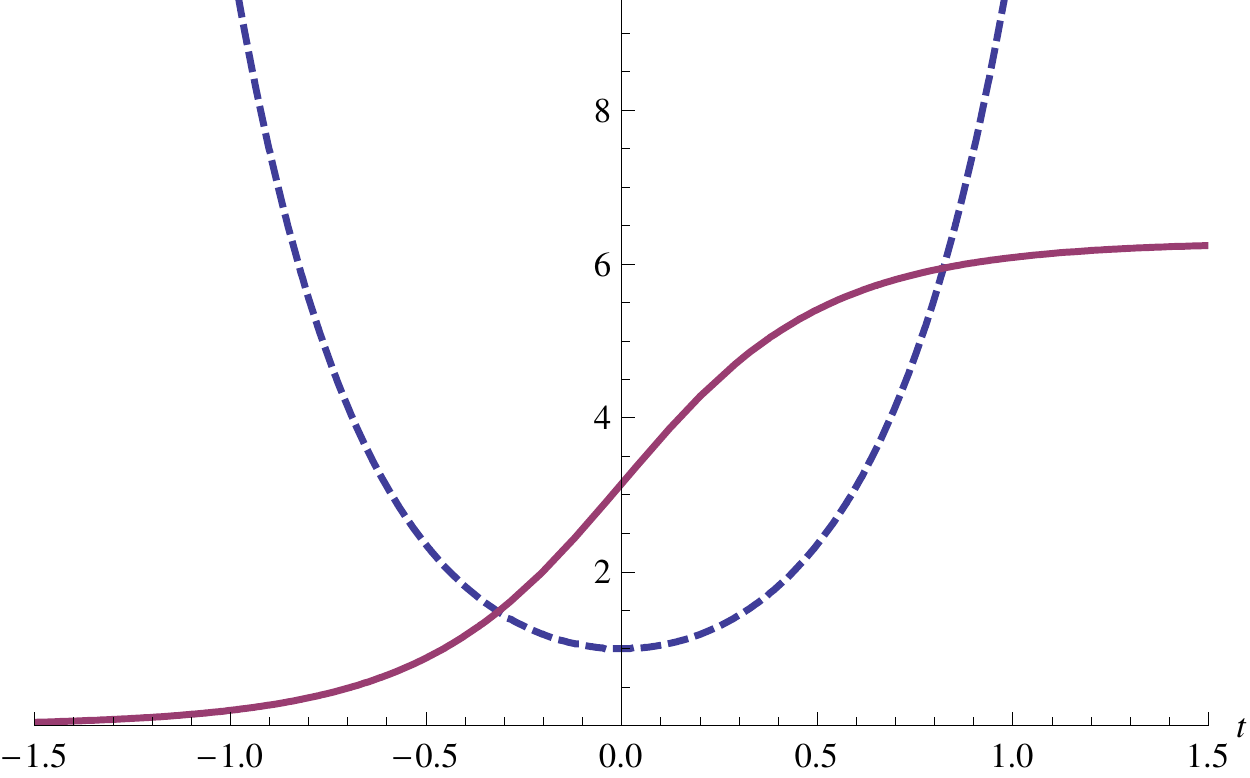}{8}{Plot of the torsion vortex solution vs. $t$. The blue dashed line is $e^{A(t)}$ while the red solid line is $\Theta(t)$. To make the plot, we have chosen $\Theta_0=0$.}

Note the curvature and torsion of this solution:
\beqn
{R^\alpha}_\beta &=& -\dot F\dot A\ {\epsilon^\alpha}_{\beta\gamma}dt\w e^\gamma- a^2 e^\alpha\w e_\beta\,,\\
{R^\alpha}_0 &=& \left(\dot h+h^2\right)dt\w e^\alpha-\frac{1}{2}\dot F\dot A\ {\epsilon^\alpha}_{\beta\gamma}e^\beta\w e^\gamma\,,\\
T^\alpha&=&- \frac{1}{2}\dot F\ {\epsilon^\alpha}_{\beta\gamma}e^\beta\w e^\gamma\,,\\
T^0&=&0\,.
\eeqn
These are non-singular for all $t\in(-\infty,\infty)$. The torsion vortex solution has divergent action, but this divergence is cancelled by boundary counterterms, the same counterterms which render the action of $AdS_4$ finite. To see this, the energy of the torsion vortex can be computed by evaluating the Euclidean action on the solution. Introducing a cutoff at $t=\pm L$, we find
\beqn
I_{tv,on-shell}&=&4a^2\int\epsilon_{\alpha\beta\gamma}dx^\alpha\w dx^\beta\w dx^\gamma\ \int dt e^{3A(t)}\\
&=&(6\int \widehat{Vol}_3)\cdot\left(\frac{4}{3} a\alpha^3 e^{3aL}+\ldots\right)\,,
\eeqn
where the ellipsis contains terms that vanish when the cutoff is removed. As in pure $AdS_4$, an appropriate counterterm is of the form \cite{Balasubramanian:1999re,Kraus:1999di}
\beq
I_{c.t.}=-\frac{4a}{3}\int_{\pa M} \epsilon_{\alpha\beta\gamma}\tilde e^\alpha\w \tilde e^\beta\w \tilde e^\gamma\,.
\eeq
In the present case, we have such a counterterm on {\it each} asymptotic boundary, and thus we find
\beq
I_{c.t.}=-2\frac{2a}{3}\alpha^3 e^{3aL}\cdot(6\int \widehat{Vol}_3)\,,
\eeq
which exactly cancels the divergent energy of the torsion vortex.

Furthermore, we note that in the Kalb-Ramond representation, the solution has
\beq
H= \dot \Theta Vol_3 =\pm 6a\alpha^3\widehat{Vol}_3\equiv \hat H \widehat{Vol}_3\,,
\eeq
where $\widehat{Vol}_3=\frac16\epsilon_{\alpha\beta\gamma}dx^\alpha\w dx^\beta\w dx^\gamma$. This corresponds to a `topological quantum number' of the kink
\beq
\int *_4H=\pm\Delta\Theta=\pm 2\pi.
\eeq

\section{The torsion vortex as the gravity dual of parity symmetry breaking}


The holographic interpretation of the torsion vortex is also of interest.
To study this, we set to zero without loss of generality the integration constant $\Theta_0=0$ and pick the plus sign in (\ref{Pf}), (\ref{Theta}). Next we need the asymptotic expansion of the vierbein which reads
\beq
\label{FGe}
\tilde{e}^\alpha =2^{-1/3}\alpha e^{\pm a(t-t_0)}\left(1+\frac{1}{3}e^{\mp 6a(t-t_0)}+\cdots\right)dx^\alpha\,\,\, {\rm for}\,\,\, t\rightarrow \pm\infty\,.
\eeq
This shows that our solution is asymptotically anti-de Sitter for both $t\rightarrow\pm\infty$. The two asymptotic AdS spaces have the same cosmological constant. From this expansion we could read the expectation value of the renormalized boundary energy momentum tensor which would be given by the coefficient of the $e^{\pm 3at}$ term  (see e.g. \cite{MPT1,MPT2}). Such a term is missing in (\ref{FGe}), hence the expectation value of the boundary energy momentum tensor is zero. 

It is not immediately apparent how to interpret these two asymptotic regimes. Are they truly distinct, or should they be identified in some way? We note that the pseudoscalar behaves in these asymptotic regimes as
\beqn
\label{Theta+t}
\Theta(t) &\rightarrow &4e^{-3a(t-t_0)}-\frac{4}{3}e^{-9a(t-t_0)}+\cdots \,\,\, {\rm for} \,\,\, t\rightarrow -\infty \,,\\
\label{Theta-t}
\Theta(t) &\rightarrow &  2\pi -4e^{3a(t-t_0)}+\frac{4}{3}e^{9a(t-t_0)}+\cdots\,\,\,{\rm for}\,\,\ t\rightarrow +\infty\,.
\eeqn
From the above we confirm that $\Theta(t)$ is dual to a dimension $\Delta=3$ boundary pseudoscalar that we denote ${\cal O}_3$. In each one of the asymptotically AdS regimes, the leading constant behavior of $\Theta(t)$ corresponds to the source (i.e., coupling constant) for ${\cal O}_3$ and the subleading term proportional to $e^{\mp 3a(t-t_0)}$ to the expectation value $\langle{\cal O}_3\rangle$. 
The two asymptotic regimes are distinguished by the behavior of $\Theta$. In fact, the essential difference is {\it parity}.

We can now describe the holography of our torsion vortex. In the $t\rightarrow -\infty$ boundary sits a three-dimensional CFT at a parity breaking vacuum state. The order parameter is the expectation value of the pseudoscalar which is $\langle {\cal O}_3\rangle =4$ in units of the AdS radius. The expectation value breaks of course the conformal invariance of the boundary theory. Then, the theory is deformed by the same pseudoscalar operator $g{\cal O}_3$ where $g$ is a marginal coupling. The torsion vortex provides the holographic description of that deformation. A solution with two asymptotic regimes is difficult to interpret in terms of the usual holographic renormalization group. Note though that in this case, at $t\rightarrow  +\infty$ the space becomes AdS with the {\it same} radius as at $t\to-\infty$. Hence, the two boundary theories have the  same `central charges'.\footnote{We use ``central charge" in $d=3$ for a quantity that counts the massless degrees of freedom at the fixed point. Such a quantity may be taken to be the coefficient in the two-point function of the energy momentum tensor or the coefficient of the free energy density. Recall that there is no conformal anomaly in $d=3$.} 

We suggest that instead of interpreting the solution in terms of an RG flow, we should think of it as a transition between two inequivalent vacua of a single theory. This statement is supported by the behavior of $\Theta(t)$ in the two asymptotic regimes.
For  $t\rightarrow \infty$ the pseudoscalar asymptotes to the configuration (\ref{Theta-t}). The interpretation is now that when the marginal coupling takes the fixed value $g_*=2\pi$ we are back to the {\it same} CFT (i.e. having the same central charge) however in a distinct parity breaking vacuum such that $\langle{\cal O}_3\rangle =-4$. In others words, the two asymptotic AdS regimes seem to describe two distinct parity breaking vacua of the same theory. The two vacua are distinguished by the expectation value of the parity breaking order parameter being $\langle {\cal O}_3\rangle =\pm 4$. Quite remarkably, we also seem to find that starting in one of the two vacua, we can reach the other by a marginal deformation with a {\it fixed} value of the deformation parameter. 

Since the marginal operator is of dimension $\Delta=3$ and parity odd, we tentatively identify it with a Chern-Simons operator of a boundary gauge field. In this case the torsion vortex induces the T-transformation in the boundary CFT \cite{Witten,LPsl2z}. In  Appendix \ref{app:C} we will argue that the three dimensional Gross-Neveu model coupled to abelian gauge fields exhibits a large-$N$ vacuum structure that matches our holographic findings. Although our bulk model is extremely simple to provide details for its possible  holographic dual, we regard this remarkable similarity as strong qualitative evidence that our torsion vortex is the gravity dual of the `tunneling' between different parity breaking vacua in three dimensions. However, in a three-dimensional quantum field theory, we do not expect that tunneling can occur because of large volume effects, and distinct vacua remain orthogonal. Thus, referring to the torsion vortex as a tunneling event should be taken figuratively. We leave to future work a more careful study of the boundary interpretation of the torsion vortex solution. An interpretation will depend on the precise topology of the boundary.\cite{uslater}

\section{Physics in the Bulk: The Superconductor Analogy}

The bulk interpretation of the exact solution is also interesting. Because the pseudoscalar field undergoes $\Theta(t)\to \Theta(t)+2\pi$ under $t$ goes from $-\infty$ to $+\infty$, the exact
solution corresponds to a topological kink. 
It satisfies 
\[ \int dt\dot \Theta=2\pi \]
In Figure 2, we plot the solution.
\myfig{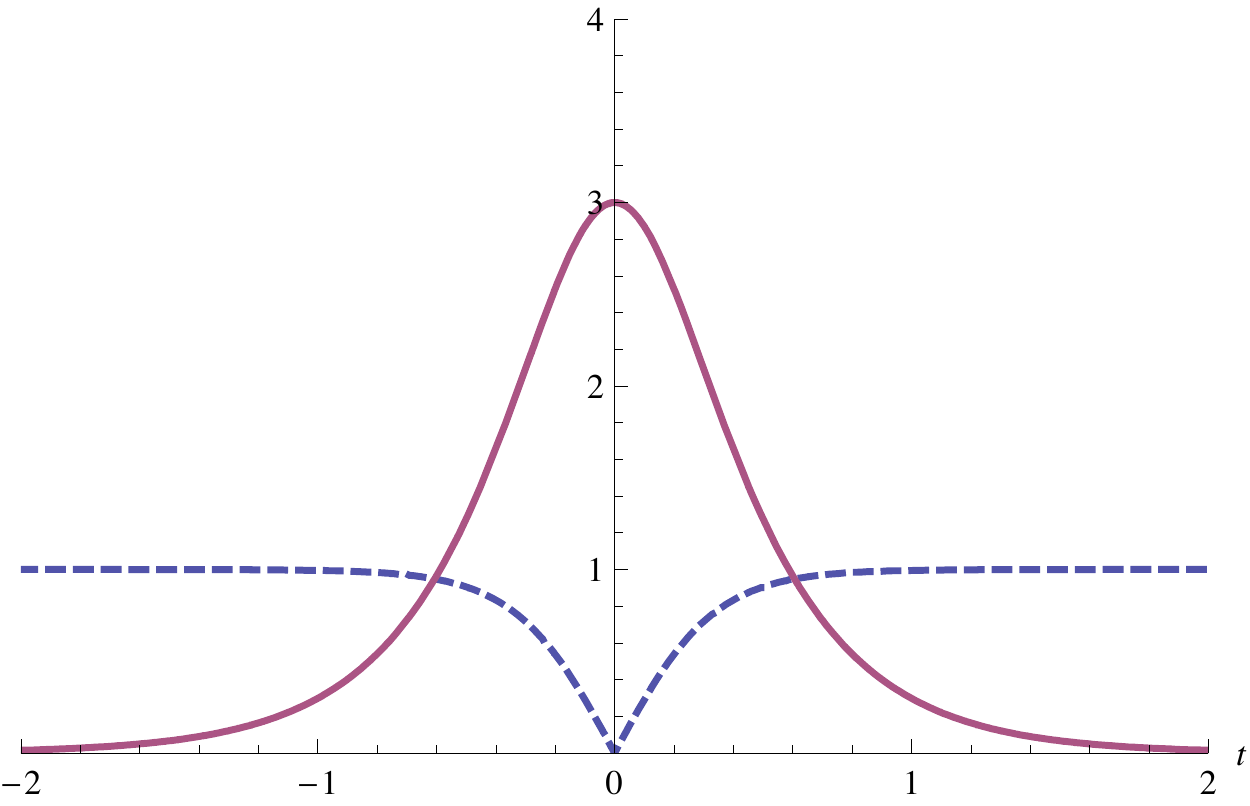}{10}{The blue dashed line is $|h(t)|$, resembling the order parameter of a superconductor, while the solid red line is $\Pi_F$, analogous to the magnetic induction of an Abrikosov vortex.}

\subsection{Gravity vortex as Abrikosov vortex}

The gravity vortex solution (\ref{eq:hdot}--\ref{Theta}) bears some resemblance to the Abrikosov vortex of superconducting systems. In this section, we will explore this and point out some possibly interesting features. The first thing to notice is that the plot in Figure \ref{fig:vortex2.pdf} is identical to the profile of an Abrikosov vortex (see for example Figure 5.1 in Ref. \cite{tinkham}.) The codimension differs,\footnote{The difference in dimensionality of the core is what we expect, since it supports a 3-form field strength in contrast to a 2-form field strength in superconductivity.} but there is a correspondence between our radial $t$-direction and the radial direction in the Abrikosov vortex, and $|h|$ and $\Pi_F$ correspond to the condensate and magnetic induction of the superconductor, respectively.
Table 1 summarizes the correspondence. 
\begin{table}[ht]
\centering
\begin{tabular}{c  c}
  \hline\hline\\
  $\h{30pt}$ \text{Abrikosov vortex}$\h{30pt}$ & $\h{30pt}$\text{Torsion vortex}$\h{30pt}$ \\[1ex]
  \hline\\
  order parameter $\Phi$ & order parameter $|h|=|{\dot A}|$\\[1ex]
    $T-T_c$  & $\Lambda$  \\[1ex]
  magnetic induction B & $\Pi_F$\\[1ex]
  magnetic field H & $\hat H$\\[1ex]
  $\ZZ$-quantized magnetic flux & $\ZZ_2$-quantized
  electric flux\\[1ex]
  \hline
\end{tabular}\label{table 1}
\caption{Abrikosov vortex v.s. Torsion vortex}
\end{table}
In this correspondence, since the order parameter is $h=\dot A$, the superconducting phase (constant order parameter) corresponds to $AdS_4$, while the normal phase corresponds to flat space ($h=0$). 
Far away from the core of the torsion vortex, the geometry is asymptotically $AdS$, but at the core the spatial slice (at $t\to t_0$) becomes flat. To see this, note that if we think of the system as a pseudoscalar coupled to torsionless gravity, the torsion vortex has ${\cc{\omega}^\alpha}_\beta=0$ and ${\cc{\omega}^\alpha}_0= \dot A\tilde e^\alpha$, and so
\beqn
{\cc{R}^\alpha}_\beta &=& - h^2 \tilde e^\alpha\w\tilde e_\beta\,,\\
{\cc{R}^\alpha}_0&=&(\dot h+h^2) dt\w\tilde e^\alpha\,,\\
\cc{T}^\alpha &=&0\,.
\eeqn
Thus, at the core, we find that the Riemann tensor has components
\begin{align}
  {R^{\alpha}}_{0\alpha 0} &\to -3 a^2 \alpha\,,\\
  {R^\alpha}_{\beta\alpha\beta} &\to 0\,.
\end{align}
This behavior is in line with an Abrikosov vortex in which there is normal phase at the core and superconducting phase away from the core. 

The analogue of the magnetic field is what we have called $\hat H$, proportional to the constant $\alpha^3$. In the vortex, the magnetic induction, analogous to $\Pi_F$, has a penetration length $\lambda\sim 1/3a$, and the coherence length of the order parameter is $\xi\sim 1/6a$. The penetration and coherence length are obtained by looking at the exponential fall-off of these quantities in the core of the vortex, away from their values in the superconducting phase.

The torsion vortex also has a quantized flux $\int *_4 H=\Delta\Theta=2\pi$.  This flux is independent of any parameters of the solution and of any rescaling of fields in the theory. Thus, this is an analogue of the quantized magnetic flux in superconductivity.

Finally, note the following interesting feature. If we take a derivative of the second equation in (\ref{eq:hdot}), we arrive at
\begin{equation}\label{dot-h-eq}
\ddot h - 6\Lambda h - 18 h^3 = 0\,.
\end{equation}
This looks like a Landau-Ginzburg equation of motion of an effective $\phi^4$ theory. This leads us to interpret $\Lambda \sim T-T_c$. Of course, there is no real temperature in the case of the torsion vortex, but we note that this implies that the penetration and coherence lengths diverge as $T\to T_c$ with exponent $1/2$, as in superconductivity.

\subsection{Multi-vortices and Vortex Condensation}

In the last section, we noted that there is a strong analogue between the torsion vortex solution and superconductivity. It is intriguing to carry the analogy further and consider multi-vortex configurations. We have noted that at the core of the torsion vortex, the spatial sections are flat. Thus, one might imagine that if it was favourable for torsion vortices to condense, as vortices do in Type I superconductors, then finite regions of normal phase (corresponding to $\Lambda=0$) would obtain. We will argue below that this can in fact occur, although the system appears not to be unstable. 

To understand the physics involved, the first step is to consider a configuration of two vortices. In the superconductivity literature, this is a standard computation. One takes two vortices separated by a distance $\ell$ and computes the Euclidean action. More precisely, we will treat this here as follows. Denoting the torsion vortex schematically as $\Phi(t_0)$, we take a configuration
\beqn
\left\{ \begin{matrix} \Phi(\ell/2),&\ t>0\cr
\Phi(-\ell/2),&\ t<0
\end{matrix}\right.\,.
\eeqn
We have taken a piecewise solution, because solutions of non-linear equations cannot be simply superimposed.
The result is not quite a solution to the equations of motion of course, failing at the midpoint between the vortices. However, if we simply evaluate the Euclidean action, we find
\beq
S_E(\ell)=4a\alpha^3 \sinh(3a\ell/2)\,.
\eeq
Note that this is positive, so one might naively conclude that the vortices repel each other. However, recall that the vortex profile exists not in flat space-time, but in the metric given by (\ref{eq:expA}), which rises asymptotically. As a result, as we move the vortices further apart, there is a corresponding rise in the metric between the vortices. So, we should directly evaluate the force
\beq
F=-\frac{dS_E}{d\ell}=-6a\alpha^3\cosh(3a\ell/2) <0\,.
\eeq 
and thus we conclude that the vortices in fact {\it attract} each other. In the superconducting analogue, this implies that we have a {\it Type I superconductor}. In such a superconductor, the number of vortices is determined by the total magnetic flux, and the vortices tend to clump together forming (potentially) finite regions of normal phase within the superconductor.

\begin{figure}[ht]
  \hfill
  \begin{minipage}[t]{.45\textwidth}
	 \includegraphics[width=8cm]{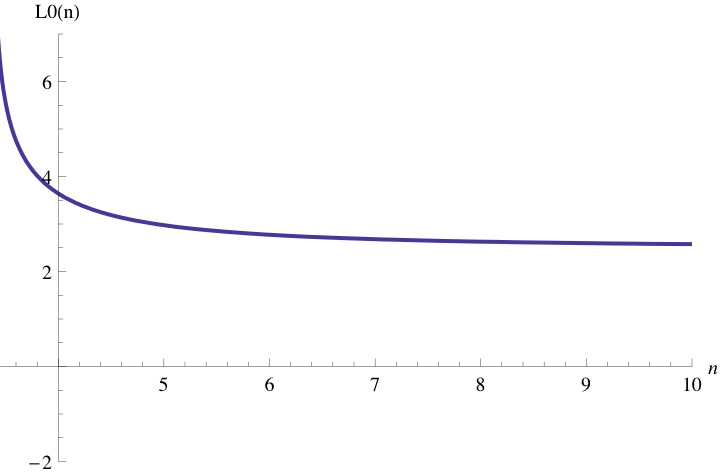}
		\caption{Size of normal state droplet vs. $n$ for multi-vortex.}
		\label{fig:mvL0.pdf}
  \end{minipage}
  \hfill
  \begin{minipage}[t]{.45\textwidth}
	 \includegraphics[width=8cm]{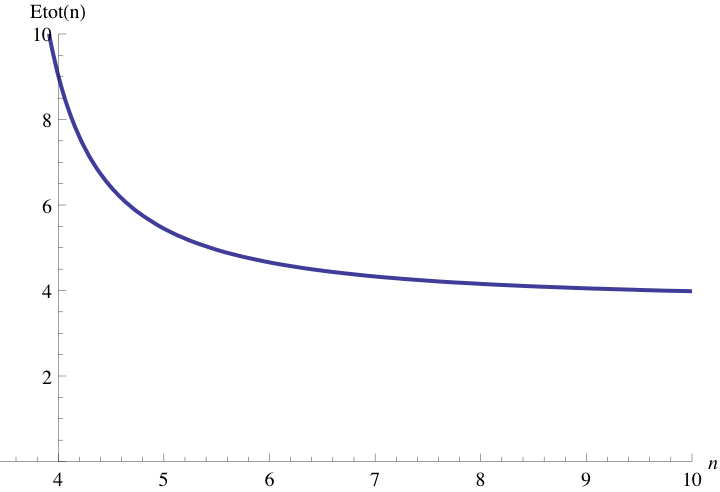}
		\caption{Energy vs. $n$ for multi-vortex.}
		\label{fig:mvEtot.pdf}
  \end{minipage}
  \hfill
\end{figure}

We now describe the analogous situation in our gravitational system. We have noted that the constant $\hat H$ plays the role of the external magnetic induction, while $H$ is the magnetic field, varying within the vortex, with $\Delta\Theta=\int *_4 H$. Following the superconducting analogue, if we put the system in a box of size $2L$ (that is we impose a cutoff on each AdS asymptotic) the flux conservation equation is of the form
\beq\label{fluxquant}
\Delta\Theta = 2L\hat H
\eeq
The vortices carry the flux in the superconductor, and so it is natural to ask what is the lowest energy configuration satisfying (\ref{fluxquant})? To analyze this, consider an array of $n$ vortices in a region of size $L_0$. We take the vortices to be equally spaced, as one can show that deviating from such a configuration causes a rise in energy. For such a configuration, the flux quantization condition (\ref{fluxquant}) gives a relation between $n$,$L_0$ and $\hat H$. Such a representative curve is shown in Fig. \ref{fig:mvL0.pdf}. 
If we solve this equation for $L_0$ as a function of $n$ and $\hat H$, we can then compute the energy as a function of $n$. One obtains a curve as in Fig. \ref{fig:mvEtot.pdf}. One notes that the energy is minimized for large $n$, and in that case, the size $L_0$ asymptotes to a fixed value, which is found to be
\beq
L_0=\frac{\hat H }{6a}\cdot 2L = \alpha^3\cdot 2L 
\eeq
We conclude that the preferred configuration, given a fixed external flux, is a continuum of vortices arrayed over a finite size region. Within this droplet, the system is in the normal phase. We have noted that the vortex core is spatially flat, and so we surmise that within the droplet, the space-time is flat. The asymptotic value of energy in Figure \ref{fig:mvEtot.pdf} is precisely minus that contributed by the cosmological constant. Again, the size of the droplet is set by the value of the external $H$-flux, and the boundary conditions are AdS. Note that for a fixed cutoff, there is a critical field (given by $\hat H=6a$) for which the entire spacetime is flat.

\section{Conclusions}

In this work we have presented in detail a simple toy model, the Nieh-Yan model, where torsion enters through the spacetime dependence of the coupling constant of the Nieh-Yan topological invariant. Although we have discussed the model directly in terms of torsion, it can classically be put into equivalent forms as either a massless pseudoscalar or a Kalb-Ramond field coupled to gravity. 
The model has an interesting and non-trivial holographic interpretation. 
In particular, we have shown that it possesses an exact bulk solution in Euclidean signature, termed the torsion vortex, having two asymptotically AdS$_4$ regimes, while the pseudoscalar acquires a kink profile. We have argued then that the holographic interpretation of this torsion vortex is a three-dimensional CFT with two distinct parity breaking vacua. Moreover, our bulk solution may imply that the deformation by a classically marginal pseudoscalar with a fixed  coupling constant induces a transition between the two parity breaking vacua separated by a domain wall, which would be at infinity in the boundary components.\cite{uslater} Remarkably, this qualitative behaviour is seen already in the three-dimensional Gross-Neveu model coupled to $U(1)$ gauge fields. The economy of our bulk model does not allow a detailed identification of the bulk and boundary theories, nevertheless  we believe that our results provide a strong base where an exact bulk/boundary dictionary for AdS$_4$/CFT$_3$ can be based. A further rather intriguing property of the torsion vortex is that it can be mapped into the standard Abrikosov vortex of superconductivity. Such a map identifies flat spacetime with a superconductor's normal phase, while AdS is identified with a superconducting phase. The cosmological constant would then measure the deviation from the `critical temperature'. A phenomenon of vortex condensation is found, similar to the analogous case in type I superconductors.

The upshot of our results is that the torsional degrees of freedom of four dimensional gravity can provide holographic descriptions for a number of interesting properties of three dimensional critical systems. It would be interesting to extend our analysis to more elaborate models where more torsional degrees of freedom become dynamical. It is also of interest to discuss whether our simple model can be embedded into M-theory. 

\subsection*{Acknowledgments}
We thank F. Larsen, D. Minic and N. Papanicolaou for discussions.  ACP is partially supported by the European RTN Program MRTN-CT-2004-512194, and RGL receives partial support from the US Department of Energy, under contract DE-FG02-91ER40709.

\appendix
\renewcommand{\theequation}{\thesection.\arabic{equation}}
\renewcommand{\thesubsection}{\thesection.\arabic{subsection}}
\setcounter{equation}{0}
\setcounter{subsection}{0}


%


\section{The $3+1$ Split}\label{app:B}

We recall the 3+1 split formalism. We will refer to the radial coordinate as $t$, although its signature
will be left arbitrary, and indicated by $\sigma_\perp=\pm 1$ ($\sigma_\perp=+1$ for $AdS_4$). The 
overall signature of the 4-manifold is denoted $\sigma=\det\eta$. In addition, we note the notation
\beqn
*_4(e^a\w e^b\w e^c)=\epsilon^{abcd}e_d\\
*_4e_d=-\frac{\sigma}{6}\epsilon_{defg}e^e\w e^f\w e^g
\eeqn
and for example
\beq
*_3\tilde e^\alpha = \frac12 {\epsilon^\alpha}_{\beta\gamma}\tilde e^\beta\wedge\tilde e^\gamma\\
\eeq
where $\tilde e^\alpha$ is defined as follows.
As
described in \cite{LP}, we split the 1-forms via
\beqn
e^0&=& Ndt\\
e^\alpha&=&\tilde e^\alpha +N^\alpha dt\\
{\omega^0}_\alpha&=&\sigma_\perp K_\alpha+{q^0}_\alpha dt\\
{\omega^\alpha}_\beta&=&\sigma{\epsilon^\alpha}_{\beta\gamma}\left( B^\gamma+Q^\gamma dt\right)
\eeqn
We then find
\beq T^\alpha =\tilde T^\alpha+dt\wedge\left\{\dot{\tilde e}^\alpha-\tilde d N^\alpha+NK^\alpha-\sigma{\epsilon^\alpha}_{\beta\gamma}Q^\beta e^\gamma-\sigma{\epsilon^\alpha}_{\beta\gamma}N^\beta B^\gamma\right\}
\eeq
\beq T^0 =\sigma_\perp K_\alpha\wedge\tilde e^\alpha+dt\wedge\left\{-\tilde d N-\sigma_\perp N_\alpha K^\alpha+{q^0}_\beta\tilde e^\beta\right\}
\eeq
and we write
\beq
{R^a}_b=\tilde R^{a}_{\kern5pt b}+dt\w r^{a}_{\kern5pt b}
\eeq
\begin{equation}\label{eq:GCtwo2}
{\tilde R^0}_{\kern5pt\alpha}=\sigma_\perp ( {\tilde {d}K}_\alpha+K_\beta\wedge\tilde\omega^\beta_{\kern5pt\alpha})\equiv\sigma_\perp (\tilde D K)_\alpha\,.
\end{equation}
\begin{equation}\label{eq:GCone1}
\tilde R^{\alpha}_{\kern5pt\beta}=^{(3)}\kern-5pt{R^{\alpha}}_{\beta}- \sigma_\perp K^\alpha\wedge K_\beta\,,
\end{equation}
with
\beq
^{(3)}R^{\alpha\beta}=
\sigma \left[\epsilon^{\alpha\beta\gamma} dB_\gamma-\sigma_\perp B^\alpha\w  B^\beta\right]
\eeq
and
\begin{eqnarray}
2\epsilon_{\alpha\beta\gamma}r^{0\alpha}\wedge {\tilde e}^\beta\wedge {\tilde e}^\gamma &=&
 2\sigma_\perp\epsilon_{\alpha\beta\gamma} \dot K^\alpha\wedge\tilde e^\beta\wedge\tilde e^\gamma+4Q_\alpha K_\beta\wedge\tilde e^\beta\wedge\tilde e^\alpha
+4q^{0\alpha}\left[\epsilon_{\alpha\beta\gamma}\tilde{T}^\beta\wedge\tilde{e}^\gamma\right]
\nonumber\end{eqnarray}
Including the Gibbons-Hawking term, which is of the form $2\sigma_\perp\int_{\pa M}\epsilon_{\alpha\beta\gamma} K^\alpha\wedge\tilde e^\beta\wedge\tilde e^\gamma$, we find
\beqn
I_{EH}&=&\int dt\wedge \Bigl\{ \dot{\tilde e}^\alpha\wedge (4\sigma_\perp\epsilon_{\alpha\beta\gamma}K^\gamma\wedge\tilde e^\beta)
+ 2N\epsilon_{\alpha\beta\gamma}\left( ^{(3)}R^{\alpha\beta}-\sigma_\perp K^\alpha\wedge K^\beta-\frac{\Lambda}{3} {\tilde e}^\alpha\wedge {\tilde e}^\beta\right)\wedge {\tilde e}^\gamma \nonumber \\
 && -4\sigma_\perp N^\alpha\epsilon_{\alpha\beta\gamma}(\tilde D K)^\beta\wedge {\tilde e}^\gamma
 -4q^{0\alpha}\epsilon_{\alpha\beta\gamma}\tilde T^\beta\w \tilde e^\gamma+4Q_\alpha\tilde e^\alpha\w K_\beta\w\tilde e^\beta\Bigl\}
\eeqn
Here, $N$ and $N^\alpha$ are the usual Lagrange multiplier fields for the lapse and shift constraints, while $q_{0\alpha}$ and $Q_\alpha$ are Lagrange multipliers that, in the pure gravity case, set the torsion and the antisymmetric part of the extrinsic curvature to zero.

\section{Parity breaking in three dimensions}\label{app:C}

Consider the three dimensional Gross-Neveu model coupled to abelian gauge fields. The Euclidean action is\footnote{We use $\bar{\psi}^i$, $\psi^i$ ($a=1,2,...,N$) two-component Dirac fermions. The $\gamma$-matrices are defined in terms of the usual Pauli matrices as $\gamma^i=\sigma^i$ $i=1,2,3$.}
\beq
\label{GNaction}
I=-\int d^3x\left[\bar{\psi}^a\left(\slash\!\!\!\partial-{\rm i}e\slash \!\!\!\!A\right)\psi^a +\frac{G}{2N}\left(\bar{\psi}^a\psi^a\right)^2+\frac{1}{4M}F_{\mu\nu}F_{\mu\nu}\right]\,.
 \eeq
$M$ is an UV mass scale.   Introducing the usual Lagrange multiplier field $\sigma$, whose equation of motion is $\sigma =\frac{-2G}{N}\bar{\psi}^a\psi^a$ we can make the action quadratic in the fermions
 \beq
 \label{GNactionsigma}
 I=-\int d^3x\left[\bar{\psi}^a\left(\slash\!\!\!\partial+\sigma-{\rm i}e\slash \!\!\!\!A\right)\psi^a -\frac{N}{2G}\sigma^2-\frac{1}{4M}F^{\mu\nu}F_{\mu\nu}\right]\,.
 \eeq
The model possesses two parity breaking vacua distinguished by the value of the pseudoscalar order parameter $\langle\sigma\rangle$. This is seen as follows: switching off the gauge fields momentarily one integrates over the fermions to produce a large-$N$ effective action as
\beq
\label{GNeffact}
{\cal Z} = \int ({\cal D}\sigma)e^{N\left[\Tr\log\left(\slash\!\!\!\partial +\sigma\right)-\frac{1}{2G}\int d^3x\sigma^2\right]}
\,.
\eeq
The path integral has a non-zero large-$N$ extremum $\sigma_*$ found by setting 
$\sigma =\sigma_* +\frac{1}{\sqrt{N}}\lambda$ 
\beq
\label{GNeffactN}
{\cal Z} =\int ({\cal D}\lambda) e^{N\left[\Tr\log\left(\slash\!\!\!\partial +\sigma_*\right)-\frac{1}{2G}\int d^3x \sigma_* +\frac{1}{\sqrt{N}}\left\{ \Tr\frac{\lambda}{\slash\!\!\!\partial+\sigma_*}-\frac{\sigma_*}{G}\int d^3x\lambda\right\}+O(1/N)\right]}
\eeq
The term in the curly brackets is the gap equation. To study it one considers a uniform momentum cutoff $\Lambda$ to obtain
\beq
\label{GNgap}
\frac{1}{G}=\int^\Lambda\frac{d^3p}{(2\pi)^3}\frac{2}{p^2+\sigma_*^2}=(\Tr 1)\left[\frac{\Lambda}{\pi^2}-\frac{|\sigma_*|}{\pi^2}\arctan\frac{\Lambda}{|\sigma_*|}\right]\,.
\eeq
Defining the critical coupling as
\beq
\label{GNcritcoupl}
\frac{1}{G_*}=\frac{\Lambda}{\pi^2}\,,
\eeq
(\ref{GNgap}) possesses a non-zero solution for $\sigma_*$ when $G>G_*$ given by
\beq
\label{sigma*}
|\sigma_*|=\frac{2\pi}{G}\left(\frac{G}{G_*}-1\right)\equiv m\,.
\eeq
The two distinct parity breaking vacua then have
\beq
\label{u}
\sigma_*=-\frac{2G}{N}\langle\bar{\psi}^a\psi^a\rangle =\pm m
\,.
\eeq
Going back to (\ref{GNactionsigma}) one can tune $G>G_*$ and start in any of the two parity breaking vacua. Suppose we start from $\sigma_*=+m$. To leading order in $N$ we have 
\beq
\label{GNsigmam}
{\cal Z}=\int ({\cal D}A_\mu)({\cal D}\bar{\psi}^a)({\cal D}\psi^a)e^{\int d^3x\left[\bar{\psi}^a\left(\slash\!\!\!\partial +m-{\rm i}e\slash\!\!\!\! A\right)\psi^a-\frac{N}{2G}m^2+O(1/\sqrt{N})-\frac{1}{4M}F^{\mu\nu}F_{\mu\nu}\right]}
\eeq
As is well known \cite{Semenoff,Redlich} for an odd number $N$ of fermions the path integral (\ref{GNsigmam}) yields an effective action for the gauge fields which for low momenta is dominated by the Chern-Simons term i.e.
\beq
\label{GNCS}
{\cal Z} \approx \int e^{S_{CS}}
\,,
\eeq
with
\beq
\label{CS}
S_{CS} ={\rm i}\frac{ke^2}{4\pi}\int d^3x \epsilon^{\mu\nu\rho}A_\mu\partial_\nu A_\rho\,,\,\,\,\,\,\,\,\, k=\frac{N}{2}
\,.
\eeq
Had we started from the $\sigma_*=-m$ vacuum, we would have found again (\ref{GNCS}), however with $k=-\frac{N}{2}$, i.e. the vacuum with $\sigma_*=-m$ yields an effective Chern-Simons action with $k=-\frac{N}{2}$.

 Consider now deforming the action (\ref{GNsigmam}) by the Chern-Simons term with a fixed coefficient as
 \beq
 \label{GNsigmamDef}
 {\cal Z}=\int ({\cal D}A_\mu)({\cal D}\bar{\psi}^i)({\cal D}\psi^i)e^{\int d^3x\left[\bar{\psi}^i\left(\slash\!\!\!\partial +m-{\rm i}e\slash\!\!\!\! A\right)\psi^i-\frac{N}{2G}m^2+O(1/\sqrt{N})-\frac{1}{4M}F^{\mu\nu}F_{\mu\nu}-{\rm i}q\int d^3x\epsilon^{\mu\nu\rho}A_\mu\partial_\nu A_\rho \right]}\,.
\eeq
If $q$ is fixed to
\beq
\label{q}
q=\frac{Ne^2}{4\pi}\,,
\eeq
the effective action for the gauge fields resulting from the fermionic path integrals in (\ref{GNsigmamDef}) is going to be {\it exactly equal}  the one obtained when we start at the  $\sigma_*=-m$ vacuum. In other words, deforming the $\sigma_*=+m$ vacuum with a Chern-Simons term with a fixed coefficient is equivalent to being in the $\sigma_*=-m$ vacuum. This is exactly analogous to the holographic interpretation of our torsion vortex.

%




%
\providecommand{\href}[2]{#2}\begingroup\raggedright\endgroup

\end{document}